\documentclass[tradiabstract]{aa}  
\usepackage{txfonts}

\usepackage{lscape}
\usepackage{supertabular}
\usepackage{longtable}

\usepackage[sort]{natbib}
\bibpunct{(}{)}{;}{a}{}{,}

\usepackage{graphicx}

\begin{document}

\title{Pulsar Timing for the {\em Fermi} Gamma-ray Space Telescope}

\author{D.~A. Smith\inst{1,2}
\and L. Guillemot\inst{1,2}
\and F. Camilo\inst{3}
\and I. Cognard\inst{4,5}
\and D. Dumora\inst{1,2} 
\and C. Espinoza\inst{7}
\and P.~C.~C. Freire\inst{8}
\and E.~V. Gotthelf\inst{3}
\and A.~K. Harding\inst{9}
\and G.~B. Hobbs\inst{10}
\and S. Johnston\inst{10}
\and V.~M. Kaspi\inst{11}
\and M. Kramer\inst{7}
\and M.~A. Livingstone\inst{11}
\and A.~G. Lyne\inst{7}
\and R.~N. Manchester\inst{10}
\and F.~E. Marshall\inst{9}
\and M.~A. McLaughlin\inst{12}
\and A. Noutsos\inst{7}  
\and S.~M. Ransom\inst{13}
\and M.~S.~E. Roberts\inst{14}
\and R.~W. Romani\inst{15}
\and B.~W. Stappers\inst{7}
\and G. Theureau\inst{4,5,6}
\and D.~J. Thompson\inst{9}
\and S.~E. Thorsett\inst{16}
\and N. Wang\inst{17}
\and P. Weltevrede\inst{10}}


\offprints{smith@cenbg.in2p3.fr}

\institute{Universit\'e de Bordeaux, Centre d'\'etudes nucl\'eaires de Bordeaux Gradignan, UMR 5797,  
 Gradignan, 33175, France
\and CNRS/IN2P3, Centre d'\'etudes nucl\'eaires de Bordeaux Gradignan, UMR 5797, Gradignan, 33175, France
\and Columbia Astrophysics Laboratory, Columbia University, New York, NY 10027, USA
\and Laboratoire de Physique et Chimie de l'Environnement, LPCE UMR 6115, CNRS/INSU, Orl\'eans, 45071, France
\and Station de radioastronomie de Nan\c cay, Observatoire de Paris, CNRS/INSU, 18330 Nan\c cay, France
\and GEPI, Observatoire de Paris, CNRS, Universit\'e Paris Diderot, Place Jules Janssen 92190 Meudon, France
\and University of Manchester, Jodrell Bank Observatory, Macclesfield, Cheshire SK11 9DL, UK
\and Arecibo Observatory, HC 3 Box 53995, Arecibo, Puerto Rico 00612, USA
\and NASA Goddard Space Flight Center, Greenbelt, MD 20771, USA
\and Australia Telescope National Facility, CSIRO, PO Box 76, Epping NSW 1710, Australia
\and McGill University, Montreal, Quebec, Canada
\and West Virginia University, Department of Physics, PO Box 6315, Morgantown, WV 26506, USA
\and National Radio Astronomy Observatory, Charlottesville, VA 22903, USA
\and Eureka Scientific, Inc., 2452 Delmer Street Suite 100, Oakland, CA 94602-3017, USA
\and Department of Physics, Stanford University, California, USA
\and Department of Astronomy \& Astrophysics, University of California, Santa Cruz, CA 95064, USA
\and National Astronomical Observatories-CAS, 40-5 South Beijing Road, Urumqi 830011, China}

\abstract{We describe a comprehensive pulsar monitoring campaign for the Large Area Telescope (LAT) on
the {\em Fermi Gamma-ray Space Telescope} (formerly GLAST). 
The detection and study of pulsars in gamma rays give insights into the populations of neutron stars 
and supernova rates in the Galaxy, into particle acceleration mechanisms in neutron star magnetospheres, 
and into the ``engines'' driving pulsar wind nebulae.
LAT's unprecedented sensitivity between 20 MeV and 300 GeV together with its $2.4$ sr 
field-of-view makes detection of many gamma-ray pulsars likely, justifying the 
monitoring of over two hundred pulsars with large spin-down powers. 
To search for gamma-ray pulsations from most of these pulsars requires a set of phase-connected timing solutions 
spanning a year or more to properly align the sparse photon arrival times. We describe 
the choice of pulsars and the instruments involved in the campaign. Attention is paid to verifications 
of the LAT pulsar software, using for example giant radio pulses from the \object{Crab} and from \object{PSR B1937+21} 
recorded at Nan\c{c}ay, and using X-ray data on \object{PSR J0218+4232} from XMM-Newton. We demonstrate 
accuracy of the pulsar phase calculations at the microsecond level\footnote{Data Table 1 is only 
available in electronic form at the CDS via anonymous ftp to cdsarc.u-strasbg.fr (130.79.128.5) 
or via http://cdsweb.u-strasbg.fr/cgi-bin/qcat?J/A+A/}.}
   
\keywords{pulsars:general -- Gamma-rays:observations -- Ephemerides }

\maketitle

%

\section{Introduction}

Forty years after the discovery of rotating neutron stars much is unknown 
about their emission processes, and in particular the radio emission mechanism is
still largely not understood \citep{DLMK04,YL08}. 
Of the nearly two thousand known pulsars, six have been detected 
in GeV gamma-rays with high confidence, using the EGRET detector on the 
{\em Compton Gamma-Ray Observatory} (CGRO) \citep{DJT99}. 

The Large Area Telescope (LAT) on the {\em Fermi} Gamma-ray Space Telescope
(formerly the {\em Gamma-ray Large Area Space Telescope}, or GLAST) went into orbit on 2008 June 11\citep{WA08}. 
The sensitivity
and time resolution of this instrument will allow it to discover tens or more of 
new gamma-ray pulsars \citep{DAS08}. Notably, it will be able to determine
the sources among the 169 unidentified EGRET sources that are pulsars.
However, even with a sensitivity more than 30 times greater than that of EGRET, the LAT's rate of gamma-ray photon detection will be small. For example, 
the \object{Crab} pulsar is the  third brightest known gamma-ray pulsar, but will trigger the 
LAT only once every 500 revolutions of the neutron star (15 seconds), on average.
While the \object{Crab} pulsar should be detected by the LAT with high confidence in less than a day, 
it will take years to detect pulsars near the sensitivity threshold,
with days separating individual photon arrival times. A search for
pulsations using gamma-ray data alone is quite difficult in these conditions 
 \citep{WA06,SR07}. Accurate knowledge of the rotation
parameters increases LAT pulsed sensitivity. However, many neutron stars 
slow down irregularly, a phenomenon known as ``timing noise'', making it difficult 
to extrapolate a pulsar's rotation frequency $\nu$ from one epoch to another. 
Consequently, in order to obtain the accurate ephemerides necessary for gamma-ray 
detection of pulsations, known pulsars must be observed regularly.

In anticipation of the {\em Fermi} launch, and mindful of the requirement for accurate, 
contemporaneous timing parameters in order to observe pulsars at gamma-ray energies, 
we began an extensive campaign of pulsar timing observations with the Parkes 64-meter 
radio telescope in Australia \citep{RM01}, the Lovell 76-meter telescope at the Jodrell Bank 
observatory near Manchester, England \citep{DM02}, and the 94-meter (equivalent) Nan\c cay radio telescope 
near Orleans, France. The Parkes telescope is the only telescope in the campaign that 
observes sources south of $-39$\degr. 
\citet{GT05} describes the 2002 FORT upgrade to the Nan\c{c}ay receiver, with the new
BON pulsar backend described in \citet{ICGT06} and \citet{FC07}. 
These observatories carry out observing programs in support 
of the {\em Fermi} mission and, between them, observe more than 200 pulsars with a large 
spin-down luminosity, $\dot E$, as described below, on a regular basis. In addition,
about 10 pulsars with weak radio emission that are particularly strong candidates 
for gamma-ray emission are being observed periodically with the Green Bank
radio telescope (GBT) and the Arecibo radio telescope. Four pulsars with no detectable
radio emission are being observed with the Rossi X-ray Timing Explorer
satellite (RXTE).  
The Urumqi Observatory \citep{WMZ01}
is using a 25 meter antenna to monitor 38 of the brighter radio pulsars.
The goal is to build a database of rotation parameters that will
allow folding of the gamma-rays as they are accumulated over the 5 to 10 year lifetime
of the LAT. This work is similar in spirit to what was done for CGRO 
(\citealp{ZA94}; \citealp{SJ95}; \citealp{DAGM96}; \citealp{KaspiThesis}).

\section{Pulsars and the Large Area Telescope}

The LAT is described by \citet{WA08}.
In brief, gamma-rays convert to electron-positron pairs in tungsten foil interleaved
with layers of silicon microstrip detectors in the tracker, yielding 
direction information. The particle cascade continues in the cesium iodide crystals
of the calorimeter, providing energy information. Scintillators
surrounding the tracker aid rejection of the charged cosmic ray background. 
The scintillators are segmented to reduce the ``backsplash'': a self-veto effect 
that reduced EGRET's sensitivity to high energy photons. 

The LAT is a 4-by-4 array of detector ``modules'' covering an area of roughly $1.7$ meters on a side. 
It is sensitive to photons with energies between $20$ MeV and $300$ GeV, 
whereas EGRET's sensitivity fell off significantly above 10 GeV.
After event reconstruction and background rejection,
the effective area for gamma-rays above 1 GeV is $> 8000$ cm$^2$ at normal incidence, as compared to
$1200$ cm$^2$ for EGRET. The angular resolution is also better than EGRET's, such that
source localisation for typical sources will be of order of $0.1$\degr \footnote{Details of the 
instrument response are maintained at 
{\mbox http://www-glast.slac.stanford.edu/software/IS/glast\_lat\_performance.htm}}.
The height-to-width aspect ratio of the LAT is $0.4$, for a field-of-view of $2.5$ sr, 
or nearly $20$\% of the sky at a given time. Combined with the large effective area, 
this makes a sky survey observation strategy possible: on a given orbit, the LAT 
will sweep the sky $35^\circ$ away from the orbital plane, covering 75\% of the sky. 
At the end of the orbit, {\em Fermi} will rock to $35^\circ$ on the other side of the orbital 
plane, and continue to scan. Thus, the entire sky is covered with good uniformity every 
three hours, and no time is lost to earth occultation. Survey mode, large effective 
area, and good localisation together give the LAT an overall steady point-source sensitivity 
30 times better than EGRET's.

Gamma-ray events recorded with the LAT have timestamps that derive from a GPS clock 
on the {\em Fermi} satellite. 
Ground tests using cosmic ray muons demonstrated that the LAT measures event times
with precision relative to UTC significantly better than a microsecond \citep{DAS08b}. 
On orbit, satellite telemetry indicates comparable accuracy.
The contribution to the barycentered time resolution from uncertainty in the LAT's position is
negligible. 

The EGRET pulsars showed a variety of pulse profiles and emission spectra and raised as many 
questions as they answered \citep{DJT04}. The high-energy emission is thought to 
arise from basic electromagnetic interactions of highly relativistic particles, namely 
synchrotron emission, curvature emission and inverse Compton emission. 
In the two main categories of models describing high-energy 
emission by pulsars, charged particles are accelerated along the 
magnetic dipole field lines by parallel electric fields. 
The ``polar cap'' model 
\citep{STU71,RS75} argues that the acceleration begins above the stellar magnetic pole, 
but can extend to the outer magnetosphere. In the ``outer gap'' model \citep{CHR86a,CHR86b} 
particles are thought to be accelerated to high energies only in the outer magnetosphere, 
in vacuum gaps between a null-charge surface and the light cylinder. 

The models predict different high-energy emission features 
such as spectra and profiles, that LAT observations may elucidate, through a 
hierarchy of observables. First, the different models have very different predictions 
of which and how many pulsars emit gamma-rays. Along with detections of radio-quiet pulsars 
in gamma-rays using blind search techniques, the LAT analysis using this timing program will constrain the ratio of radio-loud 
to radio-quiet pulsars. This ratio is different for the two emission models, with outer gap 
models predicting a much lower ratio \citep{HGG07,PG04}. Reliable flux upper limits in the 
absence of gamma-ray pulsations are useful in this context \citep{NAB96} and also require 
good timing solutions.

The second observable is the emission profile. Its shape, as the beam sweeps the Earth, 
provides a cross-section of the regions in the pulsar magnetosphere where the emission originates. 
Coupled with radio intensity and polarization profile studies, as well as absolute phase, the gamma-ray light curve 
provides information on the emission geometry, which differs significantly from one model to 
another \citep{GOB02,JCRWR94}. The EGRET pulsars typically have two peaks, 
with the first one slightly offset in phase relative to the single radio peak. Although the \object{Crab} 
pulsar breaks this trend, LAT observations will study the prevalence of this behaviour as a function 
of pulsar age or other parameters. Pulsar detections and emission profiles can 
only be achieved through solid knowledge of the pulsar's rotation and good absolute time precision. 
The timing precision will allow finely binned profiles over many years even for millisecond pulsars.

The large energy range covered by the LAT will enable measurements 
of pulsar spectral cut-offs. Although EGRET observed high-energy cut-offs in pulsar spectra 
around a few GeV, it did not have the sensitivity to measure the exact energy or shape of 
the turnovers. For instance, the LAT should provide a determination 
of the \object{Crab} pulsar's spectral cut-off energy, known only to be less than a few tens of GeV 
\citep{MT08,CELESTE02}, where EGRET lost sensitivity due to the backsplash effect. 
 The on-axis LAT energy resolution is better than 
$15$\% above 100 MeV and is better than $10$\% in the range between roughly $500$ MeV and 
$50$ GeV, and improves somewhat off-axis. The LAT should quickly measure the shape of 
the \object{Vela} pulsar spectral cut-off expected to be around 4 GeV, a powerful discriminator between polar cap 
and outer gap models and a potential diagnostic of high-energy emission altitude \citep{AH07}. 
Finally, a subset of the pulsars detected by 
the LAT will have sufficient photon numbers to allow phase-resolved spectroscopy, offering 
further insight into emission mechanisms and the beam geometry.

The LAT will monitor all pulsars continuously with a duty-cycle of roughly one-sixth, 
because of its survey mode, unlike EGRET or the \emph{Astro-rivelatore Gamma a Immagini LEggero} 
(AGILE) telescope \citep{AP04}, which went into orbit in April, 2007. A drawback of the survey 
strategy is that having the sample of gamma-photons spread over a longer duration makes phase-folding more difficult, 
as long-term timing noise may appear in pulsar spin behavior and glitches may occur \citep{SR07}. 
The need for a substantial and sustained pulsar timing campaign stems in part from this continuous 
observation, whereas pointing telescopes only require monitoring during observations of 
any given sky region. 

\section{The Timing Campaign}

\subsection{Possible Gamma-Ray Pulsars}

For a pulsar with a rotation frequency $\nu$ ($s^{-1}$) and frequency derivative $\dot \nu = {d\nu \over dt}$ (in 
units of $s^{-2}$), the spin-down power is $\dot E = - 4\pi^2 I \nu \dot \nu\ \mathrm{erg}/\mathrm{s}$ 
where the moment of inertia $I$ is taken to be $10^{45}$ g cm$^2$.
The open field-line voltage is 
$V \simeq 6.3 \times 10^{20} \sqrt{- \nu \dot \nu } \simeq 3.18 \times 10^{-3} \sqrt{\dot E }$ volts.
Above some value of $V$, or, equivalently, $\dot E$, gamma-ray emitting electron-positron 
cascades occur, with gamma-ray luminosity $L_\gamma$ increasing with $\dot E$ \citep{JA96}.
A linear dependence of $L_\gamma$ on $V$ would give 
$L_\gamma \propto -\nu^{0.5}\dot{\nu}^{0.5} \propto \sqrt{\dot{E}}$, leading to a gamma-ray 
production efficiency $\epsilon _{\gamma} = L_{\gamma} / \dot{E} \propto 1 / \sqrt{\dot{E}}$.
Analyses based on EGRET pulsar detections and upper limits have constrained 
gamma-ray luminosity laws \citep[e.g.][]{MCLC00}, an update of which yields 
 $L_\gamma \propto \sim -\nu^{-0.9}\dot{\nu}^{0.6}$. 
Empirically, although based on a small handful of gamma-ray pulsars, 
the minimum spin-down threshold seems to be near $\dot E \simeq 3 \times 10^{34}$ erg/s \citep{DJT99}.
The angular size and viewing geometry of pulsar beams is difficult to constrain and introduces
a large uncertainty in the relation between a minimum $\dot E$ and the expected gamma-ray flux. 
Bright radio pulsars may have gamma-ray beams missing the Earth's line-of-sight; 
conversely at least one bright gamma-ray pulsar, \object{Geminga}, has no detectable radio flux \citep{BFB99}. 
Balancing these issues, and keeping the list of gamma-ray pulsar candidates of reasonable 
length, we have selected pulsars with
\mbox{\boldmath   $\dot{E} > 10^{34}\ \mathrm{erg}/\mathrm{s} $  }
for LAT pulsar timing. From the ATNF online catalogue \citep{atnf} we obtain
230 such pulsars.
We give lower priority to the timing of the pulsars in globular clusters since 
they can have apparent $\dot E$ values higher than the true spin-down power 
of the neutron star, due to acceleration in the gravitational potential of the cluster.
(Notable exceptions to this are the millisecond pulsars \object{PSR B$1820-30$A} and \object{PSR B$1821-24$}.) 
This leaves us with {\bf 224 pulsars} which we believe are imperative to time regularly.

\begin{figure}
\includegraphics[width=9cm]{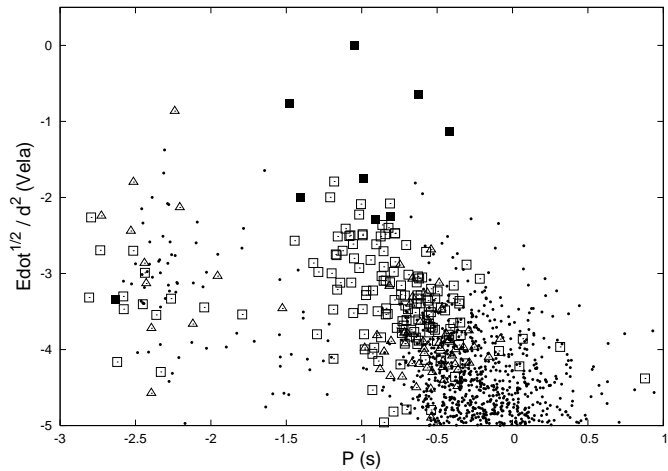}
\caption{ $\sqrt{\dot{E}}/d^{2}$, normalized to \object{Vela}, versus neutron star rotation period.
The pulsars with $\dot{E} > 3 \times 10^{34}$ erg/s (squares) and with
$10^{34}$ erg/s $< \dot{E} < 3 \times 10^{34}$ erg/s (triangles) are those
being timed regularly for the Large Area Telescope. 
The 9 solid squares correspond to the six confirmed EGRET pulsed detections
and three pulsars for which there were indications of gamma-ray pulsations. 
Dots: other pulsars. Pulsars in globular clusters are not plotted. 
Note that radio pulsars with high $\sqrt{\dot{E}}/d^{2}$ may nevertheless 
have gamma-ray beams directed away from the Earth line-of-sight and escape detection.
}\label{candidates}
\end{figure}

Table 1, available electronically,  gives the pulsar names as well as some indicators of whether they
may be gamma-ray emitters, such as $\dot E$ and associations with other high-energy 
sources\footnote{An up-to-date version is at 
https://confluence.slac.stanford.edu/\\display/GLAMCOG/Pulsars+being+timed}. The 
distance $d$ is taken from the ATNF database  (the variable ``DIST1''). It is generally based on the NE2001 model
for the Galactic distribution of free electrons \citep{NE2001} but uses other information such 
as parallax or HI absorption measurements if they are available. The uncertainty in the derived 
distances can exceed 50\%, depending on the pulsar. 
The table is sorted by decreasing $\sqrt{\dot{E}}/d^{2}$, assuming that $L_\gamma \propto V$ as
discussed above. Such a ranking ignores effects of beam geometry relative to the Earth line-of-sight, and
variations in $L_\gamma$ that may stem from, for example, the angle between the neutron star's rotation
and magnetic axes.
Figure \ref{candidates} shows $\sqrt{\dot{E}}/d^{2}$ normalized to \object{Vela}'s value versus the 
rotation period for the large $\dot E$ pulsars.

Table 1 also lists some pulsar wind nebulae (PWN) associated with young pulsars \citep{KRH04,MSER04}. Of the 
many striking results recently obtained from the HESS atmospheric Cherenkov imager array 
is the large number of Galactic sources in the TeV sky,  many of which  have been 
identified as PWN \citep[see for example][]{Aha06}. Table 1 gives TeV associations with HESS sources 
as well as a MILAGRO source \citep{Milagro}. Some of the unidentified EGRET sources 
are also likely to be PWN or pulsars. The table includes the angular distances to nearby EGRET 
$3^{rd}$ catalog sources \citep{3rdcatalog}. Many young pulsars are in or near the error boxes
for these sources, and the LAT will better localize the GeV sources, making coincidence 
tests stronger. The pulsar timing campaign will enhance searches for GeV pulsations, to 
address whether the origin is in the neutron star magnetosphere or in the nebula.
One study aimed at distinguishing between true and fortuitous associations between young pulsars 
and their PWN or EGRET counterparts predicted that $19 \pm 6$ of the EGRET-pulsar proposed associations
will be confirmed by the {\em Fermi} LAT observations \citep{MK03}. 

The table further lists those rare pulsars seen beyond radio wavelengths, either in optical 
(``O'' in the table), or in X-rays. The larger gamma-ray pulsar sample expected from the 
LAT will improve the current poor knowledge of the correlations between different types of 
high-energy emission.

Although we base the LAT timing campaign on high $\dot{E}$ pulsars, we realize that pulsar
gamma-ray emission is far from understood and therefore intend to study as many different 
pulsars as possible. The LAT's sensitivity and the continual sky-survey mode favor 
unexpected discoveries. The LAT team therefore welcomes long-term, phase-connected rotational 
ephemerides from astronomers wishing to collaborate on pulsed gamma-ray searches.

\subsection{Timing Radio-Loud Gamma-Ray Candidates}

The radio telescope time needed to monitor a given pulsar depends on the precision needed 
by the LAT, its radio flux density (e.g. $S_{1400}$ in the ATNF catalog) and pulse profile, 
and the magnitude of its timing noise. 
Simple simulations indicate that gaussian smearing of gamma-ray arrival times
barely degrades detection sensitivity, for smearing widths up to $0.05$ periods. Once detected, 
gamma-photon statistics drive the need for higher precision: the timing residuals should be smaller 
than the phase histogram bin width, which in turn should be wide enough to have at least 
several gamma-photons per bin. 

A consequence of these relatively modest timing requirements is that a given radio observation 
need only last the minimum time for detection. More crucial is the number of observations per year, which 
depends on the timing noise, correlated with $\nu$ and $\dot\nu$ and thus $\dot{E}$ \citep{CordesHelfand,ZA94}.
Gamma-ray candidates tend to be the noisiest pulsars. Illustrations of timing noise in young, high 
$\dot E$ pulsars can be found, e.g., in \citet{GEM06}. Glitches have been observed for roughly a quarter of 
the pulsars being monitored \citep{MPW08}. The bulk of the pulsars in this campaign are observed 
monthly, and a smaller number are observed weekly or bi-weekly.

The low radio fluxes of some gamma-ray pulsar candidates require long exposures 
on the biggest radio telescopes. 
We must devote time to these as they could be bright gamma-ray sources. 
Radio-faint, particularly noisy pulsars could dominate the observation schedules.

\begin{figure}
\includegraphics[width=0.48\textwidth]{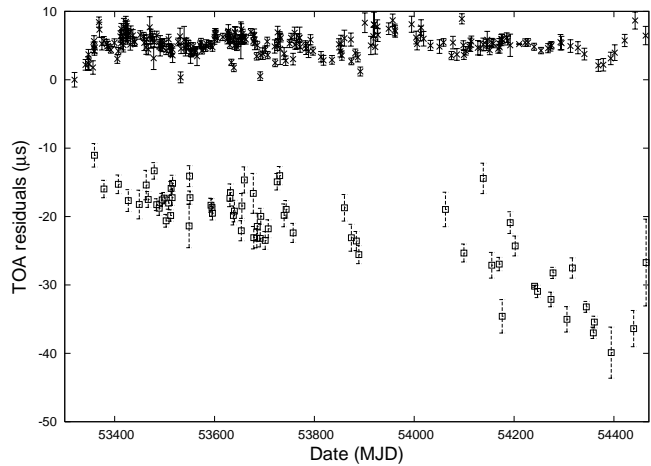}
\caption{Timing residuals for the 3.05 ms pulsar \object{B1821$-$24},
observed with the Nan\c cay radio telescope using a constant dispersion measure. 
Crosses: 1.4 GHz,  Squares: 2 GHz. Adding a dispersion measure time derivative to the 
fit aligns the residuals for the two frequencies.}
\label{residuals}
\end{figure}

Radio signals are dispersed by the interstellar medium, with a frequency dependent delay causing
signals at high radio frequencies to arrive before those at low radio frequencies. The pulsar 
Dispersion Measure (DM), or the integrated column density of free electrons along the line of sight 
from a pulsar to Earth, usually measured in cm$^{-3}$ pc, allows extrapolation of the photon arrival times from radio to infinite 
frequency, as is required for gamma-ray studies. The DM, however, can change over timescales of 
weeks to years \citep{You2007}. If the DM is inaccurate, then the reference phase $\Phi_0$ from 
the radio ephemeris (described below) will change, causing an apparent drift in the gamma-ray 
absolute phase and a smearing of the resulting gamma-ray pulse profiles. Such smearing would
compromise the multi-wavelength phase comparisons upon which beam geometry
studies are based. Therefore the timing campaign must include occasional monitoring at
multiple radio frequencies. Figure \ref{residuals} is one illustration of the magnitude of 
the dispersion for different radio frequencies. 

Another illustration of the potential effect 
of DM changes on a gamma-ray light curve is obtained using the 
DM values from the Jodrell Bank monthly \object{Crab} ephemerides.
Over the years of the Compton GRO mission (1991-1999), the 
maximum excursion in the photon time extrapolated from the radio frequency of 1400 MHz to infinite 
frequency is $0.3$ ms ($1$\%  of a rotation of the neutron star). For reference, the total DM correction from radio to gamma ray is 
$\sim 120$ ms, which is 4 \object{Crab} rotations.
For pulsars faster than the 
\object{Crab}, the effect could be larger. For most pulsars, the effect is minor. Turbulence in the 
interstellar medium also induces frequency-dependent scattering and refraction of the pulsar signal, due to 
path-length differences. Simulations show that those effects are in the order of hundreds of 
nanoseconds for observations at 1.4 GHz \citep{FC90}, and hence negligible for gamma-ray astronomy.

The radio pulsar monitoring must be sustained throughout the duration of the {\em Fermi} mission 
(i.e. for 5 to 10 years), a strain for any observatory, so other contributions are welcome.
In particular, very frequent monitoring of high $\dot{E}$, large $S_{1400}$ pulsars could allow 
significant contributions to LAT science by smaller radio telescopes.

\subsection{Radio-Quiet and Radio-Faint Pulsars}
The archetypical radio-quiet gamma-ray pulsar is \object{Geminga}, \object{PSR J0633+1746}. 
Biannual XMM satellite measurements have ensured maintenance of a phase-coherent set 
of rotation parameters over the last few years \citep{JH05}. The LAT will measure 
accurate light curves for \object{Geminga} in a few days and will maintain an accurate 
ephemeris through gamma-ray timing. The AGILE gamma-ray telescope has recently detected \object{Geminga} 
(A. Pellizzoni, private communication). 

Table 1 includes at least 15 other pulsars outside of globular clusters with $S_{1400} \leq 0.1$ mJy
(some of the others without listed $S_{1400}$ values are also faint),
requiring long radio telescope integration times, if detectable at all. 
10 of these have $\dot E>10^{36}$ erg/s, making them both especially promising gamma-ray candidates, 
and subject to especially large timing noise. 

Four high $\dot E$ pulsars are being timed with the Rossi X-ray Timing Explorer satellite (RXTE:
\object{PSR J1811$-$1925} in the center of the supernova remnant \object{G11.2$-$0.3};
the young pulsar \object{J1846$-$0258} in the core of a \object{Crab}-like pulsar wind 
nebula at the center of the bright shell-type \object{SNR Kes 75}, possibly the youngest 
known rotation-powered pulsar \citep{LKG06}; and  \object{B0540$-$69} and \object{J0537$-$6910} 
in the \object{Large Magellanic Cloud}.  

The remaining six high $\dot E$, low $S_{1400}$ pulsars are 
\object{J0205+6449}, \object{J1124$-$5916}, \object{J1747$-$2958}, \object{J1833$-$1034}, \object{J1930+1852}, and \object{J2021+3651}.
Depending on the pulse shape and the intensity of the surrounding radio nebulae, some of
these are detectable with the 70-meter class telescopes. For the others, the 
Arecibo and Green Bank (GBT) radio telescopes are more appropriate.

LAT will also perform ``blind'' searches for new radio-quiet pulsars \citep{MZ07}, that is,
search for pulsations in gamma-ray sources which are not known pulsars. Furthermore,
the LAT may detect gamma-ray sources bearing neutron star signatures, for which
no pulsations are observed, as was the case for the EGRET source \object{3EG J1835+5918} 
(\citealp{OR01}; \citealp{HCG07}). The positional uncertainty obtained with the LAT should be small enough so that
Arecibo, GBT and Parkes can perform deep radio pulsation searches.

\subsection{Public Access to Data} 
 
All public data are released through the Fermi Science Support Center (FSSC). 
The details and schedule for the LAT data releases can be found at {\mbox
http://fermi.gsfc.nasa.gov/ssc/data/policy/}.  During the first year of operations, which will be devoted
primarily to an all-sky survey, summary information about a variety of variable sources will be released.  At the
end of this period, all LAT photon data and associated science analysis tools will be released; thereafter, photon
data will be released as soon as processed, typically within days of detection.  

As LAT gamma-ray results are published, the ephemerides used will be posted on the FSSC server 
in a ``D4 FITS'' file (described below). An effort will be made to
publish a large fraction of the timing solutions acquired in this campaign around the end of the first year. In any
case, the first-year timing solutions for all 224 pulsars will be made public 6 months after the end of year 1.
Users will be asked to cite the timing parameter creators in publications, or to work with them directly.  The
intent is to update a large number of high $\dot E$ pulsar rotation ephemerides in the years following, but the 
continuation of the timing campaign will depend on the results of Cycle 1.

It is hoped  that the pulsar timing data will also be used to analyse data from instruments other than the LAT.
Atmospheric Cherenkov telescopes and neutrino detectors are just two examples of experiments that could benefit
from the timing campaign. Researchers wanting pulsar timing data may contact the authors.

\section{{\em Fermi} LAT analysis software}

\subsection{The ``Science Tools'' and the ephemerides database}

The {\em Fermi} LAT ``Science Tools'' provide a framework for analyzing gamma-ray data recorded by 
the Large Area Telescope: data selection, exposure calculation, source detection and 
identification, likelihood analysis of emission spectra, 
etc \footnote{http://fermi.gsfc.nasa.gov/ssc/data/analysis/SAE\_overview.html}.
The ``Science Tools'' are 
developed and maintained by the FSSC and instrument teams. 
This software is based on the standard \textit{ftools} developed at HEASARC 
\footnote{http://heasarc.gsfc.nasa.gov/lheasoft/ftools/ftools\_menu.html}, designed for data sets 
using the FITS format. In this section we describe \textit{gtbary} and \textit{gtpphase}, which 
are pulsar timing analysis tools.

The pulsar section of the ``Science Tools'' allows basic timing analyses within the FSSC 
framework, but is \emph{not} intended to replace specialized packages such as \emph{TEMPO} \citep{TW89} 
or \emph{TEMPO2} \citep{GLK06}. The pulsar science tools include only a subset of 
the functions provided by the those packages. 

We have tested \textit{gtbary} and \textit{gtpphase} 
with giant radio pulses from the \object{Crab} pulsar (\object{B0531+21}) and from the millisecond pulsar \object{B1937+21}, recorded 
at the Nan\c{c}ay radio telescope, X-ray photons from the binary millisecond pulsar \object{J0218+4232} 
recorded by XMM, as well as with simulated radio observations of the binary millisecond pulsar 
\object{J0437$-$4715} from Parkes.

Furthermore, we made extensive use of pulsar timing solutions (ephemerides)  
obtained from  radio or X-ray pulsar observations.
We converted ephemerides to fit the LAT format, a FITS file called ``D4'',
which contains a subset of the many parameters that pulsar astronomers provide. 
The web interface to the ephemerides database generates the D4 FITS file needed by the Science Tools.

When doing a long-term follow-up of a pulsar, one might have to use overlapping ephemerides, or 
choose between ephemerides valid on the same epoch. Those ephemerides could come from different 
observatories possibly using different analysis methods. As an example, the definition of 
the arbitrary time $T_0$ when the pulsar rotational phase equals zero, \textit{i.e.} $\Phi (T_0) = 0$ 
can differ between observatories. To ensure phase continuity when using overlapping pulsar timing 
solutions, it is important to have the template profiles used to build the ephemerides. The web-based 
tool will keep track of these template profiles. In the following we describe the pulsar timing analysis using 
the LAT software, and the different tests used to validate this process.

\subsection{Building light curves with LAT software}

Topocentric photon arrival times recorded at the observatory at finite frequency have to be 
transfered to solar system barycenter (SSB) times at infinite frequency, mainly by correcting 
times for the motion of the earth and the observatory in the solar system frame. Then one 
folds the barycenter times, using the truncated Taylor series expansion for $\Phi(t)$:

\begin{equation}
\Phi(t) = \Phi _0 + \sum _{i = 0}^{i = N} { {f_{i} \times (t - T_0)^{i+1}}\over {(i+1)!}}
\end{equation}

where $T_0$ is the reference epoch of the pulsar ephemeris, $f_i$ is the frequency derivative of order 
$i$, and $\Phi _0$ is the absolute phase, an arbitrary pulsar phase at $t = T_0$.\\

We have tested both \textit{barycenter} and \textit{phase-folding} tools. The procedure is:

\begin{itemize}
\item Conversion of the arrival times from the observatory-specific format to the LAT time format:
Mission Elapsed Time (MET) TT, which is the number of seconds since 2001 January 1 at 00:00 (UTC).
\item Calculation of the orbital or ground-based observatory position, and conversion to the LAT spacecraft position format.
\item Transfer of the topocentric times to the barycentric frame, using \textit{``gtbary''}.
\item Calculation of the pulsar phase for each arrival time, using \textit{``gtpphase''}.
\end{itemize}

\subsection{Simulated observations of an artificial pulsar}

To test the barycenter software alone, we have simulated arrival times at Nan\c{c}ay observatory and 
compared LAT barycenter software with \textit{TEMPO} and \textit{TEMPO2}.

Some time and coordinate definition differences exist between these different codes. 
Most \textit{TEMPO} pulsar timing solutions have been published using the JPL DE200 planetary ephemerides \citep{ST90}. 
\textit{TEMPO} forms barycentric times in ``Barycentric Dynamic Time'' (TDB). 
\textit{TEMPO2} uses the JPL DE405 \citep{ST98} solar system ephemerides and computes barycentric 
time in ``Barycentric Coordinate Time'' (TCB) units, taking 
into account the time dilation results from \citet{IF99}.
The LAT barycenter tool \textit{gtbary} handles both the DE200 ephemerides and the recommended
 DE405 model, also forming TDB times. The relation between TDB and TCB times is given by:

\begin{equation}
 TDB \simeq TCB - L_B \times \Delta T 
\end{equation}
where $ L_B = 1.550519767 \times 10^{-8} \pm 2 \times 10^{-17} $ and $ \Delta T = ($date $-$ 1977 
January 1, 00:00) TAI $ \times  86400$ s. TAI times refer to ``International Atomic Time''.  
(\textit{TEMPO2} has a \textit{TEMPO} emulation mode, setting the barycentric time to TDB.)
More details on time-coordinate definitions can be found in \citet{AN99}, \citet{RI01} and \citet{MP04}.

In the simulation, $10000$ arrival times are recorded on the ground, 
beginning on MJD 54100 (arbitrary), with a constant step size (no assumption of periodic  
emission is made), over 1 year. Nan\c{c}ay times are expressed in Modified Julian Days (MJD) UTC, 
at finite frequency. They first have to be moved to the LAT time format, 
at infinite frequency. The dispersion delay in the propagation of a signal at a frequency at the 
solar system barycenter $f_{SSB}$ through the interstellar medium is the following:
\begin{equation}
\Delta t = - {DM \over {K f_{SSB}^2}}
\end{equation}
where $K \equiv 2.410 \times 10^{-4}$ MHz$^{-2}$ cm$^{-3}$ pc s$^{-1}$ is the 
\textit{dispersion constant} \citep[see e.g.][]{MT77} and $DM$ is the dispersion measure.
Note that the frequency at the barycenter 
$f_{SSB}$ is different from the frequency at the observatory, due to the Doppler shift 
resulting from the motion of the observatory with respect to the pulsar \citep{ED06}. 
Higher order relativistic corrections are neglected here.
The simulated values for the pulsar position at J2000 epoch and dispersion measure 
are $(\alpha,\delta) = (20.75^{\circ},45^{\circ})$, and $DM = 0$ cm$^{-3}$ pc.

\begin{figure}
\begin{center}
\includegraphics[width=9cm]{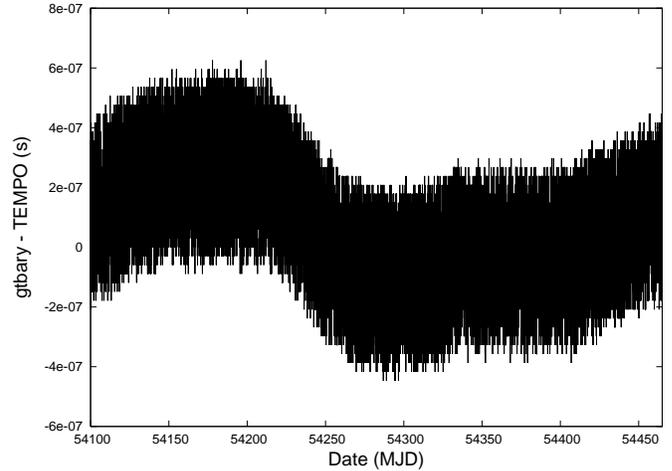}
\caption{ \textit{TEMPO}-\textit{gtbary} comparison. 
\textit{TEMPO2}-\textit{gtbary} looks very similar.}
\label{comps_fig}
\end{center}
\end{figure}

The position of the radio telescope with respect to the solar system barycenter for each 
time of arrival was calculated using the DE200 model in the \textit{TEMPO}-\textit{gtbary} 
comparison, 
and using the DE405 model with \textit{TEMPO2} in TDB mode for the 
\textit{TEMPO2}-\textit{gtbary} comparison. 
The topocentric times are then transfered to the SSB. 
The resulting differences as a function of time are shown in Figure \ref{comps_fig}. In both cases, 
time differences are below 0.7 $\mu$s, better than the instrumental precision. 
We conclude that there is agreement between the LAT barycenter code and the other standard tools.

\subsection{Giant radio pulses from the \object{Crab} pulsar and \object{B1937+21}, recorded at Nan\c{c}ay}

Giant pulses are known only from a handful of young and millisecond pulsars, 
and occupy very small windows of pulsar phase \citep[see e.g.][]{JR04,KNI06}.
Times of arrival for 3498 main component \object{Crab} GRPs with signal to noise ratio exceeding 
20 standard deviations were recorded over eight months with the Nan\c{c}ay radio telescope, as described 
in \citet{OCG08}. The radio data were de-dispersed after detection and times and observatory positions for 
each date are converted to the LAT format. Data were folded using contemporaneous Jodrell 
Bank ephemerides \citep{LPS93}, with accuracy better than 160 $\mu$s.
Event times were then converted to the barycenter using \textit{gtbary} and phase-folded with \textit{gtpphase}. 
The mean GRP arrival time is $32$ $\mu$s before that predicted by the Jodrell Bank ephemerides, 
well within ephemeris accuracies. The null phase shift of the GRPs relative to the predicted phase is 
consistent with the results of \citet{SSC03}. This demonstrates our ability to phase gamma-ray data over 
many months, even in the presence of significant timing noise, validating the codes to a few tens of $\mu$s.

\begin{figure}
\begin{center}
\includegraphics[width=9cm]{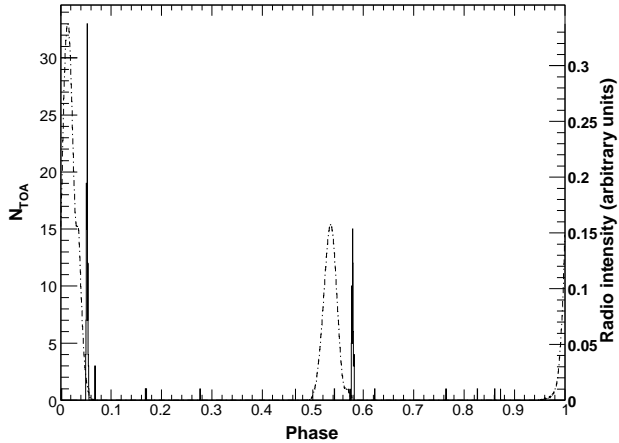}
\caption{Giant radio pulses from \object{B1937+21} recorded at Nan\c{c}ay (solid line, left-hand scale).
The $\sim$ 2 $\mu$s pulse width reflects the accuracy of the phase-folding. Also shown is the template 
radio profile at 1.4 GHz used for this study (dashed line, right-hand scale).}
\label{GRP_B1937}
\end{center}
\end{figure}

Giant pulses from \object{B1937+21} were originally discovered and studied in detail by \citet{CT96}. 
A study by \citet{KT2000} revealed that they occur in windows shorter than $10$ $\mu$s, 
$55$ to $70$ $\mu$s after the main radio pulse and interpulse, allowing
us to probe shorter timescales. Dates for 251 giant radio pulses with signal to noise exceeding 30 standard deviations were recorded 
with the Nan\c{c}ay radio telescope over three weeks.
The timing solution was derived from Nan\c{c}ay data. We phase-folded event times having corrected the 
pulsar position for proper motion. Figure \ref{GRP_B1937} shows the resulting phases, along with a pulse 
profile at 1.4 GHz. The mean delays between 
the main and secondary giant pulse components and their regular emission counterparts 
are $60.1$ and $67.3$ $\mu$s respectively, with rms deviations of $1.9$ and $2.4$ $\mu$s, consistent with Kinkhabwala 
\& Thorsett's results. The narrow pulse widths demonstrate our precision to a few $\mu$s, albeit 
for a more stable system over a shorter duration.

\subsection{X-ray data from PSR \object{J0218+4232}, observed by XMM-Newton}

Orbital movement has to be taken into account for pulsars in binary systems. The $2.3$ millisecond pulsar \object{J0218+4232} 
is in a binary system with a low mass white dwarf \citep{KU02}. It has been extensively studied at gamma-ray energies 
and is expected to be a bright {\em Fermi} source \citep{LG07}.

\begin{figure}
\begin{center}
\includegraphics[width=10cm]{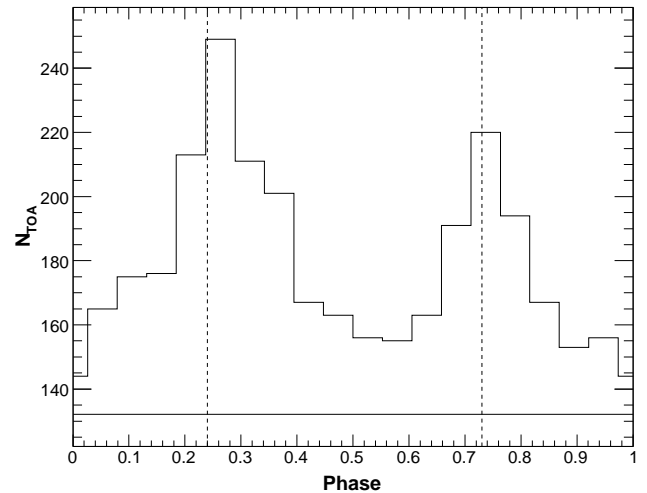}
\caption{X-ray photons from \object{J0218+4232}, recorded by the XMM-Newton satellite, between $1.6$ and $4$ keV. 
Black solid line: mean background level. Black dotted lines: peak positions in Webb et al. (2004), 
respectively $\Phi _1 = 0.24$ and $\Phi _2 = 0.73$.}
\label{XMM_J0218}
\end{center}
\end{figure}

XMM-Newton, an X-ray satellite operating between $0.1$ and $12$ keV, made a 36 ks observation of \object{J0218+4232} 
on 2002 February 11-12, with the PN camera\footnote{We thank N. Webb (CESR - Toulouse) 
for providing us with the XMM-Newton data.}. In timing mode, this instrument has a timing resolution 
of $30$ $\mu$s. Only events well calibrated in energy were retained. Pulsar data were collected using 
a rectangular region centered on the source. The background level (shown in Figure \ref{XMM_J0218}) was estimated 
by selecting data from a similar region, in the same dataset, centered about $50''$ away from the pulsar, 
where no X-ray source could be detected. Finally, only events with energy between $1.6$ and $4$ keV were selected, 
to allow comparison with studies using the same dataset done by \citet{NW04}.

Event times recorded by XMM-Newton are expressed in MET TT since 1998 January 1 at 00:00 (TT), and 
hence have to be converted to the LAT time format. 
As for the standard XMM-Newton analysis, satellite positions as a function of time were determined by a 
combination of Kepler orbital parameters and Chebyshev polynomials. Positions were interpolated 
to fit the LAT position format.

Event times were converted to the barycenter, then corrected for the pulsar orbital motion and folded, based on radio ephemerides 
given in Kuiper et al. (2002).  We tested frequencies around the nominal 
value with a $\chi ^2$ test, and found a shift in frequency of $\Delta \nu = 2.6 \times 10^{-6}$ Hz.  As noted by Webb et al. (2004), who found a similar offset, such a shift is well within the resolution of the timing data. The 
resulting phase histogram between $1.6$ and $4$ keV is shown in Figure \ref{XMM_J0218}. The peaks are 
centered on $0.26$ and $0.74$ respectively, which are within 40 $\mu$s of Webb et al's results.

\subsection{Simulated radio data for \object{PSR J0437$-$4715} observed with the Parkes telescope}

The millisecond pulsar \object{J0437$-$4715} has a pulse period of $5.76$ ms, and is in a binary system with a 
$5.74$ day orbital period. ``Post-Keplerian'' (PK) parameters, such as the rate of periastron advance, 
$\dot{\omega}$, or the rate of orbital period decay, $\dot{P_b}$, can be fit for this binary system 
\citep[e.g.][]{WVS01}. 

A $500$ day observation of \object{J0437$-$4715} at the Parkes observatory was 
simulated using \textit{TEMPO2} with the \textit{FAKE} plugin, generating 37 times of arrival.
The simulation used a timing solution for \object{J0437$-$4715} with $200$ ns accuracy, derived from real Parkes 
observations from April 1996 to March 2006 \citep{Verb08}. \textit{TEMPO2} generated a
\textit{gtpphase}-compatible solution, since the LAT Science Tools allow fewer orbital parameters than 
\textit{TEMPO2}. The accuracy of the simplified timing solution was $300$ ns.

\begin{figure}
\begin{center}
\includegraphics[width=9cm]{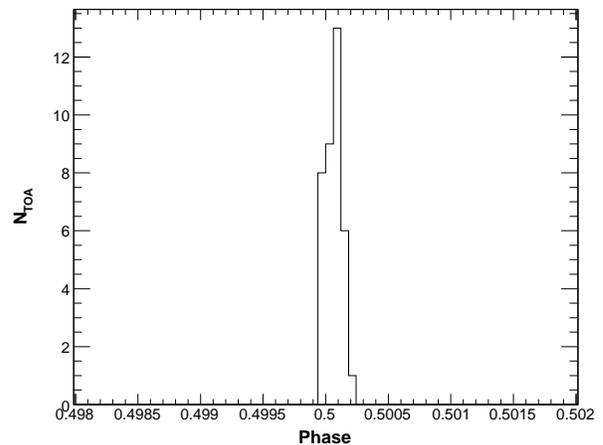}
\caption{Simulated times of arrival from the binary millisecond pulsar \object{J0437$-$4715} recorded at Parkes. 
The absolute phase is defined so that the main radio pulse is centered on 0.5.}
\label{Parkes_0437}
\end{center}
\end{figure}

Event times were transfered to the solar system barycenter, having corrected for radio dispersion and 
pulsar proper motion. Times of arrival were phase-folded based on the 
\textit{gtpphase}-compatible version of the \object{J0437$-$4715} ephemeris, yielding
the phase histogram in Figure \ref{Parkes_0437}. 
The mean phase calculated with the LAT software is delayed from the \textit{TEMPO2} mean value by $0.32$ $\mu$s, 
resulting in a validation of the code below the $\mu$s level.

\section{Conclusions}
We have motivated and described the large timing campaign that is underway for
the {\em Fermi} mission (formerly GLAST). Previous campaigns resulted in a wealth of information on young pulsars, 
and we expect this effort will expand gamma-ray pulsar detections to middle-aged, older and 
millisecond pulsars as well.  A large database of gamma-ray pulsars of many types will allow 
a study of trends and correlations in important properties such as gamma-ray flux, spectral index, 
profile shape and spectral cutoff. 

\begin{acknowledgements}
We made extensive use of the ATNF Pulsar Catalogue \citep{atnf},
http://www.atnf.csiro.au/research/pulsar/psrcat/. We also made use 
of the \object{Crab} ephemerides provided by the Jodrell Bank observatory, 
http://www.jb.man.ac.uk/pulsar/crab.html.\\

\end{acknowledgements}

\bibliographystyle{aa} 
\bibliography{article.bib}

\onecolumn
\begin{landscape}
\begin{center}
\begin{scriptsize}
\tablefirsthead{\hline
\hline
PSR	&	PSRJ	& $\frac{\sqrt{\dot{E}}}{d^2}$ & $\dot{E}$ 	& $d$  & $S_{1400}$	& $S_{400}$ & Cluster, & Optical & EGRET & EgretDist ($^\circ$) &  TeV & Notes 		\\
&		& \% of Vela 	&(erg/s)	& (kpc)	& (mJy)	& (mJy) & Galaxy & X-ray & nearby & & assoc. &	\\
\hline}
\tablehead{\multicolumn{13}{l}{\small \sl continued from previous page}\\
\hline
\hline
PSR	&	PSRJ	& $\frac{\sqrt{\dot{E}}}{d^2}$ & $\dot{E}$ 	& $d$  & $S_{1400}$	& $S_{400}$ & Cluster, & Optical & EGRET & EgretDist ($^\circ$) &  TeV & Notes 		\\
&		& \% of Vela 	&(erg/s)	& (kpc)	& (mJy)	& (mJy) & Galaxy & X-ray & nearby & & assoc. &	\\
\hline}
\bottomcaption{Pulsars being timed for the Fermi Large Area Telescope (all known pulsars with $\dot E > 10^{34}$ ergs/s), 
ordered by $\sqrt{\dot E}/d^2$, where $\dot E$ is the spin-down energy loss rate and $d$ is the distance. $\sqrt{\dot E}/d^2$ 
as an indicator of expected gamma flux suffers many large uncertainties (see text). $S_{1400}$ and $S_{400}$ are the radio 
flux intensities at $1400$ MHz and $400$ MHz, respectively.``Cluster, Galaxy'' is the name of the globular cluster or the 
host galaxy, if the pulsar is in one. ``np'' means that the source is observed in X-rays, but not pulsed. If the pulsar 
is located less than $2$\degr \ away from a $3^{rd}$ EGRET catalog source \citep{3rdcatalog}, the EGRET name and the angular 
distance are listed. The asterisk (*) indicates that the 3EG source has more than one possible counterpart in the table. 
The TeV associations are taken from \citet{HESScatalog}, available at http://www.mpi-hd.mpg.de/hfm/HESS/public/HESS\_catalog.htm, 
and from MILAGRO: \citep{Milagro}. Nearby pulsar wind nebulae (PWN) are noted in the last column \citep{KRH04,MSER04}. 
The data in the first 8 columns were obtained from the ATNF database except for the radio flux densities with the 
superscripts \textit{a}: F. Camilo, private communication, \textit{b}: \citet{JBK03}, \textit{c}: \citet{KCM98}, \textit{d}: 
\citet{RTJ96}, \textit{e}: \citet{DALM01}, \textit{f}: \citet{CLF00}. This table is 
available in electronic form at the CDS via anonymous ftp to cdsarc.u-strasbg.fr (130.79.128.5) 
or via http://cdsweb.u-strasbg.fr/cgi-bin/qcat?J/A+A/}
\tabletail{\multicolumn{13}{r}{\small \sl continuing next page\ldots}\\}
\tablelasttail{\hline}
\par
\begin{supertabular}{ccccccccccccc}
\hline
\hline
B0833$-$45 & J0835$-$4510 & 100. & 6.9e+36 & 0.3 & 1100. & 5000. &    & OX & 3EG J0834$-$4511 & 0.34 & Vela & G263.9$-$3.3, Vela X \\ 
J0633+1746 & J0633+1746 & 22. & 3.3e+34 & 0.2 & --- & --- &    & OX & 3EG J0633+1751 & 0.24 &   & G195.1+4.3, Geminga \\ 
B0531+21 & J0534+2200 & 17. & 4.6e+38 & 2.0 & 14. & 646. &    & OX & 3EG J0534+2200 & 0.13 & Crab & G184.6$-$5.8, Crab, SN1054 \\ 
J0437$-$4715 & J0437$-$4715 & 14. & 1.2e+34 & 0.2 & 142. & 550. &    &  X &   &   &   & G253.4$-$42.0 \\ 
B0656+14 & J0659+1414 & 7.42 & 3.8e+34 & 0.3 & 3.70 & 6.50 &    & OX &   &   &   & Monogem Ring \\ 
B0743$-$53 & J0745$-$5353 & 5.37 & 1.1e+34 & 0.2 & --- & 23. &    &    &   &   &   &   \\ 
J0034$-$0534 & J0034$-$0534 & 1.90 & 3.0e+34 & 0.5 & 0.61 & 17. &    &    &   &   &   &   \\ 
J0205+6449 & J0205+6449 & 1.62 & 2.7e+37 & 3.2 & 0.04 & --- &    &  X &   &   &   & G130.7+3.1, 3C 58, SN1181 \\ 
J0613$-$0200 & J0613$-$0200 & 1.58 & 1.3e+34 & 0.5 & 1.40 & 21. &    &    &   &   &   &   \\ 
J1747$-$2958 & J1747$-$2958 & 1.25 & 2.5e+36 & 2.0 & { 0.07}$^{a}$ & --- &    &    & 3EG J1744$-$3011 (*) & 0.84 & HESS J1745$-$303 & G359.23$-$0.82, Mouse \\ 
B1706$-$44 & J1709$-$4429 & 1.11 & 3.4e+36 & 2.3 & 7.30 & 25. &    &  X & 3EG J1710$-$4439 & 0.18 &   & G343.1$-$2.3 \\ 
B1055$-$52 & J1057$-$5226 & 1.06 & 3.0e+34 & 0.7 & --- & 80. &    &  X & 3EG J1058$-$5234 & 0.12 &   &   \\ 
J1740+1000 & J1740+1000 & 0.99 & 2.3e+35 & 1.2 & 9.20 & 3.10 &    &    &   &   &   &   \\ 
B1951+32 & J1952+3252 & 0.98 & 3.7e+36 & 2.5 & 1.00 & 7.00 &    &    &   &   &   & G69.0+2.7, CTB 80 \\ 
J1357$-$6429 & J1357$-$6429 & 0.90 & 3.1e+36 & 2.5 & 0.44 & --- &    &  X &   &   &   &   \\ 
J1833$-$1034 & J1833$-$1034 & 0.84 & 3.4e+37 & 4.7 & 0.07 & --- &    &    &   &   & HESS J1833$-$105 & G21.5$-$0.9 \\ 
B1509$-$58 & J1513$-$5908 & 0.76 & 1.8e+37 & 4.2 & 0.94 & 1.50 &    &  X &   &   & HESS J1514$-$591 & G320.4$-$1.2, MSH 15-52 \\ 
B1257+12 & J1300+1240 & 0.74 & 1.9e+34 & 0.8 & 2.00 & 20. &    &  np &   &   &   &   \\ 
J1524$-$5625 & J1524$-$5625 & 0.73 & 3.2e+36 & 2.8 & 0.83 & --- &    &    &   &   &   &   \\ 
J1531$-$5610 & J1531$-$5610 & 0.69 & 9.1e+35 & 2.1 & 0.60 & --- &    &    &   &   &   &   \\ 
B1046$-$58 & J1048$-$5832 & 0.60 & 2.0e+36 & 2.7 & 6.50 & --- &    &    & 3EG J1048$-$5840 & 0.14 &   & G287.4+0.58 \\ 
B0355+54 & J0358+5413 & 0.56 & 4.5e+34 & 1.1 & 23. & 46. &    &    &   &   &   &   \\ 
J0940$-$5428 & J0940$-$5428 & 0.50 & 1.9e+36 & 3.0 & 0.66 & --- &    &    &   &   &   &   \\ 
J1930+1852 & J1930+1852 & 0.44 & 1.2e+37 & 5.0 & 0.06 & --- &    &  X & 3EG J1928+1733 & 1.46 &   & G54.1+0.3 \\ 
B1259$-$63 & J1302$-$6350 & 0.37 & 8.2e+35 & 2.8 & 1.70 & --- &    &  np &   &   & HESS J1302$-$638 &   \\ 
J0834$-$4159 & J0834$-$4159 & 0.36 & 9.9e+34 & 1.7 & 0.19 & --- &    &    &   &   &   &   \\ 
J1909$-$3744 & J1909$-$3744 & 0.36 & 2.2e+34 & 1.1 & --- & { 3.}$^{b}$ &    &    &   &   &   &   \\ 
B0906$-$49 & J0908$-$4913 & 0.35 & 4.9e+35 & 2.5 & 10. & 28. &    &    &   &   &   & G270.3$-$1.0 \\ 
J1509$-$5850 & J1509$-$5850 & 0.34 & 5.2e+35 & 2.6 & 0.15 & --- &    &    &   &   &   &   \\ 
B1823$-$13 & J1826$-$1334 & 0.34 & 2.8e+36 & 3.9 & 2.10 & --- &    &  np & 3EG J1826$-$1302 (*) & 0.55 & HESS J1825$-$137 & G18.0$-$0.7 \\ 
J1809$-$1917 & J1809$-$1917 & 0.34 & 1.8e+36 & 3.5 & 2.50 & --- &    &  np &   &   & HESS J1809$-$193 &   \\ 
J0538+2817 & J0538+2817 & 0.32 & 4.9e+34 & 1.5 & 1.90 & 8.20 &    &  X &   &   &   & S147 \\ 
J1811$-$1925 & J1811$-$1925 & 0.32 & 6.4e+36 & 5.0 & --- & --- &    &  X &   &   & HESS J1809$-$193 & G11.2$-$0.3, SN 386 \\ 
J1420$-$6048 & J1420$-$6048 & 0.31 & 1.0e+37 & 5.6 & 0.90 & --- &    &  X & 3EG J1420$-$6038 & 0.17 & HESS J1420$-$607 & G313.6+0.3, Kookaburra \\ 
B1800$-$21 & J1803$-$2137 & 0.31 & 2.2e+36 & 3.9 & 7.60 & 23. &    &  np &   &   & HESS J1804$-$216 &   \\ 
J1046+0304 & J1046+0304 & 0.30 & 1.4e+34 & 1.1 & 0.30 & --- &    &    &   &   &   &   \\ 
B0114+58 & J0117+5914 & 0.30 & 2.2e+35 & 2.2 & 0.30 & 7.60 &    &    &   &   &   &   \\ 
J2229+6114 & J2229+6114 & 0.28 & 2.2e+37 & 7.2 & 0.25 & --- &    &  X & 3EG J2227+6122 & 0.54 &   & G106.6+3.1 \\ 
J1718$-$3825 & J1718$-$3825 & 0.28 & 1.2e+36 & 3.6 & 1.30 & --- &    &    & 3EG J1714$-$3857 & 1.18 & HESS J1718$-$385 &   \\ 
B1727$-$33 & J1730$-$3350 & 0.27 & 1.2e+36 & 3.5 & 3.20 & 9.20 &    &    & 3EG J1734$-$3232 & 1.57 &   &   \\ 
B0740$-$28 & J0742$-$2822 & 0.27 & 1.4e+35 & 2.1 & 15. & 296. &    &    &   &   &   &   \\ 
J1617$-$5055 & J1617$-$5055 & 0.27 & 1.6e+37 & 6.8 & { 0.5}$^{c}$ & --- &    &  X &   &   & HESS J1616$-$508 &   \\ 
J1843$-$1113 & J1843$-$1113 & 0.27 & 6.0e+34 & 1.7 & 0.10 & --- &    &    &   &   &   &   \\ 
J2129$-$5721 & J2129$-$5721 & 0.26 & 2.3e+34 & 1.4 & 1.40 & 14. &    &    &   &   &   &   \\ 
J1124$-$5916 & J1124$-$5916 & 0.26 & 1.2e+37 & 6.5 & 0.08 & --- &    &  X &   &   &   & G292.0+1.8, MSH 11$-$54 \\ 
J1846$-$0258 & J1846$-$0258 & 0.25 & 8.1e+36 & 6.0 & --- & --- &    &  X &   &   & HESS J1846$-$029 & G29.7$-$0.3, Kes 75 \\ 
J1913+1011 & J1913+1011 & 0.24 & 2.9e+36 & 4.8 & 0.50 & --- &    &    &   &   & HESS J1912+101 &   \\ 
J1911$-$1114 & J1911$-$1114 & 0.23 & 1.2e+34 & 1.2 & 0.50 & 31. &    &    & 3EG J1904$-$1124 & 1.96 &   &   \\ 
J2043+2740 & J2043+2740 & 0.23 & 5.6e+34 & 1.8 & --- & {15.}$^{d}$ &    &  X &   &   &   &   \\ 
J0855$-$4644 & J0855$-$4644 & 0.22 & 1.1e+36 & 3.9 & 0.20 & --- &    &    &   &   &   &   \\ 
J0218+4232 & J0218+4232 & 0.22 & 2.4e+35 & 2.7 & 0.90 & 35. &    &  X & 3EG J0222+4253 & 1.03 &   &   \\ 
J1739$-$3023 & J1739$-$3023 & 0.20 & 3.0e+35 & 2.9 & 1.00 & --- &    &    & 3EG J1744$-$3011 (*) & 1.10 &   &   \\ 
J1831$-$0952 & J1831$-$0952 & 0.20 & 1.1e+36 & 4.0 & 0.33 & --- &    &    &   &   &   &   \\ 
B1957+20 & J1959+2048 & 0.20 & 1.6e+35 & 2.5 & 0.40 & 20. &    &  np &   &   &   & G59.2$-$4.7 \\ 
J1105$-$6107 & J1105$-$6107 & 0.20 & 2.5e+36 & 5.0 & 0.75 & --- &    &  np & 3EG J1102$-$6103 & 0.86 &   & MSH 11$-$62 \\ 
B1821$-$24 & J1824$-$2452 & 0.19 & 2.2e+36 & 4.9 & 0.18 & 40. & M28  &  X &   &   &   &   \\ 
B1853+01 & J1856+0113 & 0.19 & 4.3e+35 & 3.3 & 0.19 & 3.40 &    &    & 3EG J1856+0114 & 0.05 &   & G34.7$-$0.4, W44, 3C 392 \\ 
B1757$-$24 & J1801$-$2451 & 0.18 & 2.6e+36 & 5.2 & 0.85 & 7.80 &    &  np & 3EG J1800$-$2338 & 1.26 &   & G5.27$-$0.9,G5.4-1.2? \\ 
B0611+22 & J0614+2229 & 0.18 & 6.2e+34 & 2.1 & 2.20 & 29. &    &    & 3EG J0617+2238 & 0.69 &   &   \\ 
B1719$-$37 & J1722$-$3712 & 0.16 & 3.3e+34 & 1.9 & 3.20 & 25. &    &    &   &   &   &   \\ 
J1835$-$1106 & J1835$-$1106 & 0.16 & 1.8e+35 & 2.8 & 2.20 & 30. &    &    &   &   &   &   \\ 
B0540+23 & J0543+2329 & 0.15 & 4.1e+34 & 2.1 & 9.00 & 29. &    &    &   &   &   &   \\ 
J1913+0904 & J1913+0904 & 0.14 & 1.6e+35 & 3.0 & 0.07 & --- &    &    &   &   &   &   \\ 
J0857$-$4424 & J0857$-$4424 & 0.13 & 2.6e+34 & 1.9 & 0.88 & 12. &    &    &   &   &   &   \\ 
J0729$-$1448 & J0729$-$1448 & 0.13 & 2.8e+35 & 3.5 & 0.70 & --- &    &    &   &   &   &   \\ 
B1737$-$30 & J1740$-$3015 & 0.12 & 8.2e+34 & 2.7 & 6.40 & 25. &    &    & 3EG J1744$-$3011 (*) & 0.86 &   &   \\ 
J1928+1746 & J1928+1746 & 0.11 & 1.6e+36 & 5.8 & 0.25 & --- &    &    & 3EG J1928+1733 & 0.29 &   &   \\ 
J1015$-$5719 & J1015$-$5719 & 0.11 & 8.3e+35 & 5.1 & 0.90 & --- &    &    & 3EG J1014$-$5705 & 0.47 &   &   \\ 
B1830$-$08 & J1833$-$0827 & 0.11 & 5.8e+35 & 4.7 & 3.60 & --- &    &    &   &   & HESS J1834$-$087 &   \\ 
J1837$-$0604 & J1837$-$0604 & 0.11 & 2.0e+36 & 6.4 & 0.70 & --- &    &    & 3EG J1837$-$0606 (*) & 0.18 &   &   \\ 
J1740$-$5340A & J1740$-$5340A & 0.10 & 1.4e+35 & 3.4 & { 1.}$^{e}$ & --- & NGC6397  &    &   &   &   &   \\ 
B1449$-$64 & J1453$-$6413 & 0.10 & 1.9e+34 & 2.1 & 14. & 230. &    &    &   &   &   &   \\ 
J1637$-$4642 & J1637$-$4642 & 0.099 & 6.4e+35 & 5.1 & 0.78 & --- &    &    & 3EG J1639$-$4702 & 0.55 &   &   \\ 
J0631+1036 & J0631+1036 & 0.098 & 1.7e+35 & 3.7 & 0.80 & 1.50 &    &  X &   &   &   &   \\ 
J1702$-$4310 & J1702$-$4310 & 0.096 & 6.3e+35 & 5.1 & 0.72 & --- &    &    &   &   &   &   \\ 
J1301$-$6305 & J1301$-$6305 & 0.094 & 1.7e+36 & 6.7 & 0.46 & --- &    &    &   &   & HESS J1301$-$631 &   \\ 
J1828$-$1101 & J1828$-$1101 & 0.092 & 1.6e+36 & 6.6 & 2.90 & --- &    &    &   &   &   &   \\ 
B1620$-$26 & J1623$-$2631 & 0.091 & 2.0e+34 & 2.2 & 1.60 & 15. & M4  &    & 3EG J1626$-$2519 & 1.35 &   &   \\ 
B1634$-$45 & J1637$-$4553 & 0.087 & 7.5e+34 & 3.2 & 1.10 & 15. &    &    & 3EG J1639$-$4702 & 1.17 &   &   \\ 
J1702$-$4128 & J1702$-$4128 & 0.082 & 3.4e+35 & 4.8 & 1.10 & --- &    &    &   &   & HESS J1702$-$420 &   \\ 
J1705$-$3950 & J1705$-$3950 & 0.081 & 7.4e+34 & 3.3 & 1.50 & --- &    &    &   &   &   &   \\ 
J1016$-$5857 & J1016$-$5857 & 0.080 & 2.6e+36 & 8.0 & 0.46 & --- &    &    & 3EG J1013$-$5915 (*) & 0.89 &   & G284.3$-$1.8 \\ 
J0901$-$4624 & J0901$-$4624 & 0.080 & 4.0e+34 & 2.8 & 0.46 & --- &    &    &   &   &   &   \\ 
B2334+61 & J2337+6151 & 0.080 & 6.2e+34 & 3.1 & 1.40 & 10. &    &  np &   &   &   &   \\ 
B1317$-$53 & J1320$-$5359 & 0.076 & 1.7e+34 & 2.3 & --- & 18. &    &    & 3EG J1316$-$5244 & 1.75 &   &   \\ 
J1841$-$0345 & J1841$-$0345 & 0.073 & 2.7e+35 & 4.8 & 1.40 & --- &    &    & 3EG J1837$-$0423 & 1.29 &   &   \\ 
J1830$-$0131 & J1830$-$0131 & 0.069 & 2.3e+34 & 2.6 & 0.35 & --- &    &    &   &   &   &   \\ 
J1119$-$6127 & J1119$-$6127 & 0.068 & 2.3e+36 & 8.4 & 0.80 & --- &    &  np &   &   &   & G292.2$-$0.5 \\ 
J1549$-$4848 & J1549$-$4848 & 0.068 & 2.3e+34 & 2.7 & 0.47 & 17. &    &    &   &   &   &   \\ 
J1816$-$0755 & J1816$-$0755 & 0.065 & 2.5e+34 & 2.8 & 0.17 & --- &    &    &   &   &   &   \\ 
J1857+0143 & J1857+0143 & 0.064 & 4.5e+35 & 5.8 & 0.74 & --- &    &    & 3EG J1856+0114 & 0.63 & HESS J1858+020 &   \\ 
B1001$-$47 & J1003$-$4747 & 0.064 & 3.0e+34 & 2.9 & --- & 6.00 &    &    &   &   &   &   \\ 
B1610$-$50 & J1614$-$5048 & 0.064 & 1.6e+36 & 7.9 & 2.40 & --- &    &    &   &   &   &   \\ 
B1828$-$11 & J1830$-$1059 & 0.060 & 3.6e+34 & 3.2 & 1.40 & 2.10 &    &    &   &   &   &   \\ 
J1648$-$4611 & J1648$-$4611 & 0.059 & 2.1e+35 & 5.0 & 0.58 & --- &    &    & 3EG J1655$-$4554 & 1.68 &   &   \\ 
B1643$-$43 & J1646$-$4346 & 0.057 & 3.6e+35 & 5.8 & 0.98 & --- &    &    &   &   &   & G341.2+0.9 \\ 
J1906+0746 & J1906+0746 & 0.057 & 2.7e+35 & 5.4 & 0.55 & 0.90 &    &    &   &   &   &   \\ 
B0136+57 & J0139+5814 & 0.055 & 2.1e+34 & 2.9 & 4.60 & 28. &    &    &   &   &   &   \\ 
J1738$-$2955 & J1738$-$2955 & 0.050 & 3.7e+34 & 3.5 & 0.29 & --- &    &    & 3EG J1736$-$2908 (*) & 1.07 &   &   \\ 
B1730$-$37 & J1733$-$3716 & 0.050 & 1.5e+34 & 2.8 & 3.40 & --- &    &    &   &   &   &   \\ 
B0021$-$72F & J0024$-$7204F & 0.049 & 1.4e+35 & 4.9 & { 0.15}$^{f}$ & --- & 47Tuc  &    &   &   &   &   \\ 
J1723$-$3659 & J1723$-$3659 & 0.049 & 3.8e+34 & 3.5 & 1.50 & --- &    &    &   &   &   &   \\ 
J1715$-$3903 & J1715$-$3903 & 0.049 & 6.9e+34 & 4.1 & 0.46 & --- &    &    & 3EG J1714$-$3857 & 0.33 &   &   \\ 
J1601$-$5335 & J1601$-$5335 & 0.048 & 1.0e+35 & 4.5 & 0.22 & --- &    &    &   &   &   &   \\ 
J1514$-$5925 & J1514$-$5925 & 0.048 & 3.5e+34 & 3.5 & 0.27 & --- &    &    &   &   & HESS J1514$-$591 &   \\ 
B1937+21 & J1939+2134 & 0.048 & 1.1e+36 & 8.3 & 10. & 240. &    &  X &   &   &   &   \\ 
B1820$-$30A & J1823$-$3021A & 0.046 & 8.3e+35 & 7.9 & 0.72 & 16. & NGC6624  &    &   &   &   &   \\ 
J1828$-$1057 & J1828$-$1057 & 0.046 & 5.5e+34 & 4.0 & 0.23 & --- &    &    &   &   &   &   \\ 
J0905$-$5127 & J0905$-$5127 & 0.045 & 2.4e+34 & 3.3 & 1.10 & 12. &    &    &   &   &   &   \\ 
J1112$-$6103 & J1112$-$6103 & 0.045 & 4.5e+36 & 12.2 & 1.40 & --- &    &    &   &   &   &   \\ 
B1221$-$63 & J1224$-$6407 & 0.044 & 1.9e+34 & 3.1 & 3.90 & 48. &    &    &   &   &   &   \\ 
B1727$-$47 & J1731$-$4744 & 0.042 & 1.1e+34 & 2.8 & 12. & 190. &    &    &   &   &   &   \\ 
B1356$-$60 & J1359$-$6038 & 0.041 & 1.2e+35 & 5.2 & 7.60 & 105. &    &    &   &   &   &   \\ 
J1921+0812 & J1921+0812 & 0.040 & 2.3e+34 & 3.5 & 0.66 & --- &    &    &   &   &   &   \\ 
B1915+13 & J1917+1353 & 0.039 & 3.8e+34 & 4.0 & 1.90 & 43. &    &    &   &   &   &   \\ 
B0021$-$72E & J0024$-$7204E & 0.039 & 8.8e+34 & 4.9 & { 0.21}$^{f}$ & --- & 47Tuc  &    &   &   &   &   \\ 
J1838$-$0549 & J1838$-$0549 & 0.039 & 1.0e+35 & 5.1 & 0.29 & --- &    &    & 3EG J1837$-$0606 (*) & 0.50 & HESS J1841$-$055 &   \\ 
J1548$-$5607 & J1548$-$5607 & 0.039 & 8.5e+34 & 4.9 & 1.00 & --- &    &    &   &   &   &   \\ 
J1835$-$0944 & J1835$-$0944 & 0.039 & 5.6e+34 & 4.4 & 0.41 & --- &    &    &   &   &   &   \\ 
 J2021+3651  & J2021+3651 & 0.038 & 3.4e+36 & 12.4 & 0.10 & --- &    &  X & 3EG J2021+3716 (*) & 0.41 & MGRO J2019+37 & G75.2+0.1 \\ 
B1535$-$56 & J1539$-$5626 & 0.037 & 1.3e+34 & 3.1 & 4.60 & --- &    &    &   &   &   &   \\ 
J1638$-$4608 & J1638$-$4608 & 0.036 & 9.4e+34 & 5.2 & 0.33 & --- &    &    & 3EG J1639$-$4702 & 0.91 &   &   \\ 
J1910$-$5959D & J1910$-$5959D & 0.036 & 5.2e+34 & 4.5 & 0.05 & --- & NGC6752  &    &   &   &   &   \\ 
J1841$-$0524 & J1841$-$0524 & 0.035 & 1.0e+35 & 5.3 & 0.20 & --- &    &    & 3EG J1837$-$0606 (*) & 1.39 & HESS J1841$-$055 &   \\ 
J0024$-$7204O & J0024$-$7204O & 0.033 & 6.5e+34 & 4.9 & { 0.10}$^{f}$ & --- & 47Tuc  &    &   &   &   &   \\ 
B1754$-$24 & J1757$-$2421 & 0.033 & 4.0e+34 & 4.4 & 3.90 & 20. &    &    & 3EG J1800$-$2338 & 0.97 &   &   \\ 
J1452$-$5851 & J1452$-$5851 & 0.032 & 3.5e+34 & 4.3 & 0.24 & --- &    &    &   &   &   &   \\ 
J1541$-$5535 & J1541$-$5535 & 0.032 & 1.1e+35 & 5.7 & 0.22 & --- &    &    &   &   &   &   \\ 
J1643$-$4505 & J1643$-$4505 & 0.031 & 9.4e+34 & 5.6 & 0.28 & --- &    &    &   &   &   &   \\ 
B1607$-$52 & J1611$-$5209 & 0.031 & 3.4e+34 & 4.3 & 1.20 & --- &    &    &   &   &   &   \\ 
B1338$-$62 & J1341$-$6220 & 0.030 & 1.4e+36 & 11.1 & 1.90 & --- &    &  X &   &   &   & G308.8$-$0.1 \\ 
J1907+0918 & J1907+0918 & 0.029 & 3.2e+35 & 7.8 & 0.29 & 0.40 &    &    &   &   &   &   \\ 
J1016$-$5819 & J1016$-$5819 & 0.029 & 4.1e+34 & 4.7 & 0.31 & --- &    &    & 3EG J1013$-$5915 (*) & 1.23 &   &   \\ 
J0537$-$6910 & J0537$-$6910 & 0.029 & 4.9e+38 & 49.4 & --- & --- &  LMC &  X & 3EG J0533$-$6916 & 1.20 &   & G279.6$-$31.7, N 157B \\ 
B1930+22 & J1932+2220 & 0.028 & 7.5e+35 & 9.8 & 1.20 & 7.80 &    &    &   &   &   &   \\ 
J1853$-$0004 & J1853$-$0004 & 0.028 & 2.1e+35 & 7.2 & 0.87 & --- &    &    & 3EG J1856+0114 & 1.46 &   &   \\ 
J0024$-$7204U & J0024$-$7204U & 0.028 & 4.6e+34 & 4.9 & { 0.06}$^{f}$ & --- & 47Tuc  &    &   &   &   &   \\ 
J1019$-$5749 & J1019$-$5749 & 0.028 & 1.8e+35 & 6.9 & 0.80 & --- &    &    & 3EG J1014$-$5705 (*) & 1.64 &   &   \\ 
B1556$-$57 & J1600$-$5751 & 0.027 & 1.1e+34 & 3.5 & 1.40 & 20. &    &    &   &   &   &   \\ 
J1841+0130 & J1841+0130 & 0.027 & 1.2e+34 & 3.6 & 0.06 & --- &    &    &   &   &   &   \\ 
J1543$-$5459 & J1543$-$5459 & 0.026 & 3.8e+34 & 4.8 & 0.62 & --- &    &    &   &   &   &   \\ 
J0954$-$5430 & J0954$-$5430 & 0.026 & 1.6e+34 & 3.9 & 0.36 & --- &    &    &   &   &   &   \\ 
J1737$-$3137 & J1737$-$3137 & 0.026 & 6.0e+34 & 5.5 & 0.80 & --- &    &    & 3EG J1734$-$3232 & 1.19 &   &   \\ 
J1650$-$4921 & J1650$-$4921 & 0.025 & 1.9e+34 & 4.1 & 0.16 & --- &    &    &   &   &   &   \\ 
J1815$-$1738 & J1815$-$1738 & 0.025 & 3.9e+35 & 8.8 & 0.25 & --- &    &    &   &   & HESS J1813$-$178 &   \\ 
B1822$-$14 & J1825$-$1446 & 0.025 & 4.1e+34 & 5.1 & 2.60 & --- &    &    & 3EG J1824$-$1514 (*) & 0.52 & HESS J1826$-$148  &   \\ 
B1821$-$19 & J1824$-$1945 & 0.024 & 3.0e+34 & 4.7 & 4.90 & 71. &    &    &   &   &   &   \\ 
J1853+0056 & J1853+0056 & 0.024 & 4.0e+34 & 5.1 & 0.21 & --- &    &    & 3EG J1856+0114 & 0.68 &   &   \\ 
J1115$-$6052 & J1115$-$6052 & 0.023 & 1.6e+34 & 4.1 & 0.38 & --- &    &    &   &   &   &   \\ 
J1903+0925 & J1903+0925 & 0.023 & 3.2e+34 & 4.9 & 0.20 & --- &    &    &   &   &   &   \\ 
J1413$-$6141 & J1413$-$6141 & 0.023 & 5.6e+35 & 10.1 & 0.61 & --- &    &    & 3EG J1410$-$6147 & 0.80 &   &   \\ 
J1756$-$2225 & J1756$-$2225 & 0.023 & 3.1e+34 & 5.0 & 0.25 & --- &    &    & 3EG J1800$-$2338 & 1.50 &   &   \\ 
J1406$-$6121 & J1406$-$6121 & 0.022 & 2.2e+35 & 8.2 & 0.36 & --- &    &    & 3EG J1410$-$6147 & 0.90 &   &   \\ 
J0024$-$7204T & J0024$-$7204T & 0.021 & 2.7e+34 & 4.9 & --- & --- & 47Tuc  &    &   &   &   &   \\ 
B1718$-$35 & J1721$-$3532 & 0.021 & 4.5e+34 & 5.6 & 11. & --- &    &    &   &   &   &   \\ 
B1508$-$57 & J1512$-$5759 & 0.021 & 1.3e+35 & 7.3 & 6.00 & --- &    &    &   &   &   &   \\ 
B1832$-$06 & J1835$-$0643 & 0.019 & 5.6e+34 & 6.2 & 1.30 & --- &    &    & 3EG J1837$-$0606 & 0.78 &   &   \\ 
B1838$-$04 & J1841$-$0425 & 0.019 & 3.9e+34 & 5.7 & 2.60 & 2.60 &    &    & 3EG J1837$-$0423 (*) & 1.02 &   &   \\ 
J1538$-$5551 & J1538$-$5551 & 0.019 & 1.1e+35 & 7.5 & 0.25 & --- &    &    &   &   &   &   \\ 
J0024$-$7204Q & J0024$-$7204Q & 0.018 & 2.0e+34 & 4.9 & { 0.05}$^{f}$ & --- & 47Tuc  &    &   &   &   &   \\ 
J1138$-$6207 & J1138$-$6207 & 0.018 & 3.0e+35 & 9.7 & 0.49 & --- &    &    &   &   &   &   \\ 
J1156$-$5707 & J1156$-$5707 & 0.018 & 4.3e+34 & 6.0 & 0.19 & --- &    &    &   &   &   &   \\ 
J1734$-$3333 & J1734$-$3333 & 0.018 & 5.6e+34 & 6.5 & 0.50 & --- &    &    & 3EG J1734$-$3232 & 1.03 &   &   \\ 
J1412$-$6145 & J1412$-$6145 & 0.018 & 1.2e+35 & 7.8 & 0.47 & --- &    &    & 3EG J1410$-$6147 & 0.53 &   &   \\ 
J1743$-$3153 & J1743$-$3153 & 0.017 & 5.8e+34 & 6.6 & 0.50 & --- &    &    & 3EG J1744$-$3011 & 1.71 &   &   \\ 
J1632$-$4757 & J1632$-$4757 & 0.017 & 5.0e+34 & 6.4 & 0.30 & --- &    &    & 3EG J1639$-$4702 & 1.92 & HESS J1632$-$478 &   \\ 
J1123$-$6259 & J1123$-$6259 & 0.017 & 1.0e+34 & 4.3 & 0.56 & 11. &    &    &   &   &   &   \\ 
J1909+0912 & J1909+0912 & 0.017 & 1.3e+35 & 8.2 & 0.35 & --- &    &    &   &   &   &   \\ 
J1650$-$4502 & J1650$-$4502 & 0.017 & 1.1e+34 & 4.4 & 0.35 & --- &    &    & 3EG J1655$-$4554 & 1.41 &   &   \\ 
J1638$-$4417 & J1638$-$4417 & 0.016 & 3.9e+34 & 6.2 & 0.21 & --- &    &    &   &   &   &   \\ 
B0540$-$69 & J0540$-$6919 & 0.016 & 1.5e+38 & 49.4 & 0.02 & --- &  LMC & OX & 3EG J0533$-$6916 & 1.80 &   & G279.7$-$31.5, N158A \\ 
J1834$-$0731 & J1834$-$0731 & 0.015 & 1.7e+34 & 5.1 & 1.00 & --- &    &    & 3EG J1837$-$0606 & 1.57 &   &   \\ 
J1052$-$5954 & J1052$-$5954 & 0.015 & 1.3e+35 & 8.5 & 0.15 & --- &    &    & 3EG J1048$-$5840 & 1.70 &   &   \\ 
B2148+52 & J2150+5247 & 0.015 & 1.1e+34 & 4.6 & 2.00 & 16. &    &    &   &   &   &   \\ 
J1649$-$4653 & J1649$-$4653 & 0.015 & 1.1e+34 & 4.7 & 0.31 & --- &    &    & 3EG J1655$-$4554 & 1.71 &   &   \\ 
J1551$-$5310 & J1551$-$5310 & 0.015 & 8.3e+34 & 7.8 & 0.54 & --- &    &    &   &   &   &   \\ 
J1843$-$0702 & J1843$-$0702 & 0.015 & 1.2e+34 & 4.8 & 0.17 & --- &    &    & 3EG J1837$-$0606 & 1.85 &   &   \\ 
J1907+0919 & J1907+0919 & 0.014 & 2.2e+34 & 5.8 & --- & --- &    &    &   &   &   &   \\ 
J1349$-$6130 & J1349$-$6130 & 0.014 & 1.2e+34 & 5.0 & 0.58 & --- &    &    &   &   &   &   \\ 
J1913+0832 & J1913+0832 & 0.014 & 7.4e+34 & 7.9 & 0.60 & --- &    &    &   &   &   &   \\ 
J1838$-$0453 & J1838$-$0453 & 0.013 & 8.3e+34 & 8.1 & 0.33 & --- &    &    & 3EG J1837$-$0423 (*) & 0.59 &   &   \\ 
J1627$-$4706 & J1627$-$4706 & 0.013 & 2.5e+34 & 6.1 & 0.10 & --- &    &    &   &   &   &   \\ 
J1837$-$0559 & J1837$-$0559 & 0.013 & 1.6e+34 & 5.4 & 0.50 & --- &    &    & 3EG J1837$-$0606 (*) & 0.15 & HESS J1841$-$055 &   \\ 
J1248$-$6344 & J1248$-$6344 & 0.013 & 8.6e+34 & 8.3 & 0.12 & --- &    &    &   &   &   &   \\ 
J1845$-$0743 & J1845$-$0743 & 0.013 & 1.3e+34 & 5.2 & 2.70 & --- &    &    &   &   &   &   \\ 
J1839$-$0905 & J1839$-$0905 & 0.012 & 1.4e+34 & 5.4 & 0.16 & --- &    &    &   &   &   &   \\ 
J1845$-$0316 & J1845$-$0316 & 0.012 & 3.9e+34 & 7.0 & 0.35 & --- &    &    &   &   & HESS J1846$-$029 &   \\ 
J1801$-$2154 & J1801$-$2154 & 0.012 & 1.2e+34 & 5.3 & 0.18 & --- &    &    & 3EG J1800$-$2338 & 1.75 &   &   \\ 
J1853+0545 & J1853+0545 & 0.012 & 1.2e+34 & 5.3 & 1.60 & --- &    &    &   &   &   &   \\ 
J1907+0731 & J1907+0731 & 0.011 & 1.5e+34 & 5.7 & 0.35 & --- &    &    &   &   &   &   \\ 
B1557$-$50 & J1600$-$5044 & 0.011 & 2.8e+34 & 6.7 & 17. & --- &    &    &   &   &   &   \\ 
J1755$-$2534 & J1755$-$2534 & 0.011 & 3.5e+34 & 7.1 & 0.17 & --- &    &    &   &   &   &   \\ 
J1452$-$6036 & J1452$-$6036 & 0.011 & 1.5e+34 & 5.8 & 1.40 & --- &    &    &   &   &   &   \\ 
J1839$-$0321 & J1839$-$0321 & 0.011 & 3.6e+34 & 7.2 & 0.27 & --- &    &    & 3EG J1837$-$0423 & 1.22 &   &   \\ 
J1632$-$4818 & J1632$-$4818 & 0.011 & 4.8e+34 & 7.8 & 0.39 & --- &    &    &   &   & HESS J1632$-$478 &   \\ 
B1841$-$05 & J1844$-$0538 & 0.011 & 2.3e+34 & 6.5 & 2.20 & --- &    &    & 3EG J1837$-$0606 & 1.83 &   &   \\ 
J1550$-$5418 & J1550$-$5418 & 0.011 & 1.0e+35 & 9.6 & 3.30 & --- &    &    &   &   &   &   \\ 
B2000+32 & J2002+3217 & 0.010 & 1.2e+34 & 5.7 & 1.20 & 5.50 &    &    &   &   &   &   \\ 
J1907+0345 & J1907+0345 & 0.0094 & 2.3e+34 & 7.2 & 0.17 & --- &    &    &   &   &   &   \\ 
J1842$-$0905 & J1842$-$0905 & 0.0091 & 1.0e+34 & 5.9 & 0.81 & --- &    &    &   &   &   &   \\ 
J1946+2611 & J1946+2611 & 0.0089 & 1.1e+34 & 6.1 & --- & 1.50 &    &    &   &   &   &   \\ 
J1726$-$3530 & J1726$-$3530 & 0.0084 & 3.5e+34 & 8.4 & 0.30 & --- &    &    &   &   &   &   \\ 
B1636$-$47 & J1640$-$4715 & 0.0083 & 1.2e+34 & 6.5 & 1.20 & --- &    &    & 3EG J1639$-$4702 & 0.38 &   &   \\ 
J1904+0800 & J1904+0800 & 0.0082 & 3.7e+34 & 8.6 & 0.36 & --- &    &    &   &   &   &   \\ 
J1820$-$1529 & J1820$-$1529 & 0.0080 & 4.0e+34 & 8.9 & 0.61 & --- &    &    & 3EG J1824$-$1514 & 0.87 &   &   \\ 
B2011+38 & J2013+3845 & 0.0076 & 2.9e+34 & 8.4 & 6.40 & 26. &    &    & 3EG J2016+3657 & 1.95 &   &   \\ 
J1908+0909 & J1908+0909 & 0.0075 & 3.6e+34 & 9.0 & 0.22 & --- &    &    &   &   &   &   \\ 
B1855+02 & J1857+0212 & 0.0074 & 2.2e+34 & 8.0 & 1.60 & --- &    &    & 3EG J1856+0114 & 1.07 & HESS J1858+020 &   \\ 
J1903+0601 & J1903+0601 & 0.0073 & 1.5e+34 & 7.2 & 0.26 & --- &    &    & 3EG J1903+0550 & 0.21 &   &   \\ 
J1515$-$5720 & J1515$-$5720 & 0.0072 & 1.0e+34 & 6.6 & 0.20 & --- &    &    &   &   &   &   \\ 
J1853+0011 & J1853+0011 & 0.0070 & 2.1e+34 & 8.1 & 0.30 & --- &    &    & 3EG J1856+0114 & 1.21 &   &   \\ 
J1342+2822B & J1342+2822B & 0.0068 & 5.4e+34 & 10.4 & 0.01 & --- & M3  &    &   &   &   &   \\ 
J1806$-$2125 & J1806$-$2125 & 0.0068 & 4.3e+34 & 9.8 & 1.10 & --- &    &    &   &   &   &   \\ 
J1626$-$4807 & J1626$-$4807 & 0.0066 & 2.7e+34 & 8.9 & 0.37 & --- &    &    &   &   &   &   \\ 
J1020$-$6026 & J1020$-$6026 & 0.0065 & 9.6e+34 & 12.3 & 0.14 & --- &    &    &   &   &   &   \\ 
J1305$-$6203 & J1305$-$6203 & 0.0055 & 1.6e+34 & 8.5 & 0.62 & --- &    &    & 3EG J1308$-$6112 & 1.08 &   &   \\ 
J1843$-$0355 & J1843$-$0355 & 0.0055 & 1.8e+34 & 8.8 & 0.80 & --- &    &    & 3EG J1837$-$0423 & 1.59 &   &   \\ 
J1735$-$3258 & J1735$-$3258 & 0.0054 & 2.4e+34 & 9.6 & 0.46 & --- &    &    & 3EG J1734$-$3232 & 0.66 &   &   \\ 
B2127+11E & J2129+1210E & 0.0050 & 7.0e+34 & 12.9 & --- & 0.20 & M15  &    &   &   &   &   \\ 
B1758$-$23 & J1801$-$2304 & 0.0050 & 6.2e+34 & 12.6 & 2.20 & --- &    &    & 3EG J1800$-$2338 (*) & 0.65 &   &   \\ 
J1043$-$6116 & J1043$-$6116 & 0.0046 & 1.7e+34 & 9.5 & 0.91 & --- &    &    &   &   &   &   \\ 
J1908+0839 & J1908+0839 & 0.0045 & 1.5e+34 & 9.3 & 0.49 & --- &    &    &   &   &   &   \\ 
J1808$-$2024 & J1808$-$2024 & 0.0041 & 5.0e+34 & 13.1 & --- & --- &    &    &   &   &   &   \\ 
J1812$-$1910 & J1812$-$1910 & 0.0035 & 1.9e+34 & 11.2 & 0.22 & --- &    &    &   &   &   &   \\ 
J1327$-$6400 & J1327$-$6400 & 0.0031 & 5.6e+34 & 15.5 & 0.36 & --- &    &    &   &   &   &   \\ 
B2127+11F & J2129+1210F & 0.0026 & 1.9e+34 & 12.9 & --- & 0.10 & M15  &    &   &   &   &   \\ 
J1524$-$5706 & J1524$-$5706 & 0.0024 & 1.0e+34 & 11.4 & 0.41 & --- &    &    &   &   &   &   \\ 
J1216$-$6223 & J1216$-$6223 & 0.0013 & 1.3e+34 & 16.6 & 0.15 & --- &    &    &   &   &   &   \\ 
J0535$-$6935 & J0535$-$6935 & 0.00031 & 5.6e+34 & 49.4 & 0.05 & --- &  LMC &    & 3EG J0533$-$6916 & 0.59 &   &   \\ 
B0456$-$69 & J0455$-$6951 & 0.00014 & 1.2e+34 & 49.4 & --- & 0.60 &  LMC &    &   &   &   &   \\ 
\end{supertabular}
\end{scriptsize}
\end{center}
\label{tab1}
\end{landscape}

\end{document}